\documentclass[12pt,preprint]{aastex}
\usepackage{graphicx}
\shorttitle{Starbursts in galaxy mergers}
\shortauthors{Teyssier, Chapon and Bournaud}
\begin{document}
\title{The driving mechanism of starbursts in galaxy mergers}
\author{Romain Teyssier$^{1,2}$, Damien Chapon$^{1}$ \& Fr\'ed\'eric Bournaud$^{1}$}
\affil{$^1$Laboratoire AIM Paris-Saclay, CEA/IRFU/SAp, CNRS, Universit\'ee Paris Diderot, 91191 Gif-sur-Yvette Cedex, France.\\
$^2$Institute for Theoretical Physics, University of Z\"urich, CH-8057 Z\"urich, Switzerland}

\begin{abstract}
We present hydrodynamic simulations of a major merger of disk galaxies, and study the ISM dynamics and star formation properties. High spatial and mass resolutions of 12~pc and $4\times10^4$~M$_{\sun}$ allow to resolve cold and turbulent gas clouds embedded in a warmer diffuse phase. We compare to lower resolution models, where the multiphase ISM is not resolved and is modeled as a relatively homogeneous and stable medium. While merger-driven bursts of star formation are generally attributed to large-scale gas inflows towards the nuclear regions, we show that once a realistic ISM is resolved, the dominant process is actually gas fragmentation into massive and dense clouds and rapid star formation therein. As a consequence, star formation is more efficient by a factor of up to $\sim 10$ and is also somewhat more extended, while the gas density probability distribution function (PDF) rapidly evolves towards very high densities. We thus propose that the actual mechanism of starburst triggering in galaxy collisions can only be captured at high spatial resolution and when the cooling of gas is modeled down to less than $10^3$~K. Not only does our model reproduce the properties of the Antennae system, but it also explains the ``starburst mode'' revealed recently in high-redshift mergers compared to quiescent disks.
\end{abstract}


\section{Introduction}

Galaxies formed a significant fraction of their stars during violent interactions and mergers, as probed by observations \citep[e.g.,][]{elbaz03} and numerical models \citep[e.g.,][]{Hopkins:2006p64}. The strongest bursts of star formation are produced by major mergers of equal-mass galaxies \citep{barnes, mh94, cox06}, while the star formation (SF) activity rapidly decreases with increasing mass ratio \citep{cox08, dimat08, manthey08,knierman09}. Nevertheless, the overall contribution of mergers to the SF budget of galaxies remains uncertain and quiescent star formation in isolated systems could dominate \citep{jogee09,robaina09,fs09,Daddi:2010p4750}. In general, merger-induced SF is assumed to be centrally-concentrated, and to fuel the formation of bulges and compact spheroids rather than extended disks \citep{naab03,bournaud04}.

The mechanism usually invoked to explain merger-induced starbursts is that the interaction with a companion induces an asymetry in the gas response, such as strong spiral arms and extended tidal tails, and the gas subsequently undergoes gravity torques. Inside the corotation radius, usually a few kpc, these torques are negative and the gas flows inwards: the density rapidly increases in the central regions, and so does the star formation rate. The mechanism was first detailed by \cite{barnes}, and the resulting SFR were quantified in simulations by, e.g., \cite{mh94} and \cite{cox06}: the SFR in a merging pair can increase by a factor of a few tens compared to the same galaxies in isolation, but the increase is generally much more modest -- typically a factor of a few units compared to isolated disks \citep{dimat08}.

A fundamental and general observational fact about interacting galaxies is that their star formation proceeds in giant molecular clouds and star clusters, which can be 10-100 times more massive than in normal spiral galaxies, leading in particular to the formation of Super Star Clusters \citep[SSCs;][]{Whitmore:2007p4470}. A theoretical explanation is proposed to be an increased gas turbulence \citep{elmegreen-pairs, struck-pairs}, so that the Jeans mass, which sets the typical mass of gas clouds, becomes larger while the free-fall time of these gas clouds decreases. This mechanism of massive gas clouds formation could trigger the SF activity of interacting galaxies, independently of the traditional central inflow mechanism. 

However, most existing models of galaxy mergers, in particular those studying the SF activity, do not resolve clustered star formation in dense cold gas clouds: star formation is instead treated as a relatively smooth process (at least at scales of 100-1000~pc), taking place in a relatively homogeneous ISM supported by thermal pressure instead of a cloudy ISM supported by turbulent motions. Star cluster formation is then only indirectly modeled using sub-grid recipes \citep[e.g.,][]{li}. This is caused by a limited spatial and mass resolution, and/or the absence of model for gas cooling below $\sim 10^4$~K in many cases. Only a few models of galaxy mergers can directly resolve cold gas clouds and clustered SF \citep[e.g.,][]{wetzstein07, bournaud08} but the properties of merger-induced SF were not studied.
  
In this paper, we present adaptive mesh refinement (AMR) hydrodynamic simulations of a major galaxy merger. Our models have a maximal spatial resolution of 12~pc, and a refinement strategy ensuring that gas can cool down to a few 100~K. Gas fragmentation into dense clouds and star formation therein can thus be directly captured, at least down to masses of $10^6$ M$_\odot$. A pressure floor avoids artificial fragmentation. A realistic multiphase ISM with dense clouds embedded in a warmer phase naturally arises in similar models of disk galaxies \citep{Tasker:2006p2660, Agertz:2009p355, Kim:2009p4605}.

Using this model for a merger of two galaxies with the interaction orbit of the Antennae galaxies, we study the interaction-induced SF properties and compare to lower-resolution simulations with a smoother, warmer ISM. We show that the process of gas fragmentation into massive clouds and rapid star formation therein dominates the merger-induced activity, while gas inflows become less efficient when a clumpy multiphase ISM is modelled. These results suggest that merger-driven SF does not follow the processes revealed in lower resolution simulations. Consequences include a potentially stronger starburst, but also a more extended distribution of gas and SF. With these new properties of merger-induced SF, our model can explain the properties of the Antennae galaxies. We also propose an interpretation for the different SF efficiencies observed in quiescent disks and active mergers \citep{Daddi:2010p4354, Genzel:2010p4738}.

\section{Model and parameters}

\subsection{Gas physics and star formation model}
We use the AMR code RAMSES to evolve the dark matter and the stellar component using a Particle Mesh solver, and the gas component using a second-order Godunov scheme \citep{Teyssier:2002p451}. Throughout our study, we model star formation with a Schmidt law: the local star formation rate is $\dot \rho_* = \epsilon_* \rho_{\rm gas}/t_{\rm ff}$, where $t_{\rm ff} = \sqrt{3\pi / 32G\rho_{\rm gas}}$ is the free-fall time computed at the gas density $\rho_{\rm gas}$. The efficiency of star formation is controlled by the parameter $\epsilon_* \simeq 1$~\%. Star formation occurs only in dense enough regions (molecular clouds), defined by the gas density $n_{\rm H}$ being greater than some threshold value $n_{*}$. These two main parameters are usually calibrated using observations of nearby galaxies and the so--called Kennicutt-Schmidt (KS) law. The real efficiency is high with a high threshold, but models with a limited resolution have to use a low threshold combined with a low efficiency \citep{Wada:2007p2426}, which globally reproduces the same KS law \citep{Elmegreen:2002p2535}.

\begin{figure*}
\centering
\vspace{-0.3in}
\includegraphics[angle=90,width=7in]{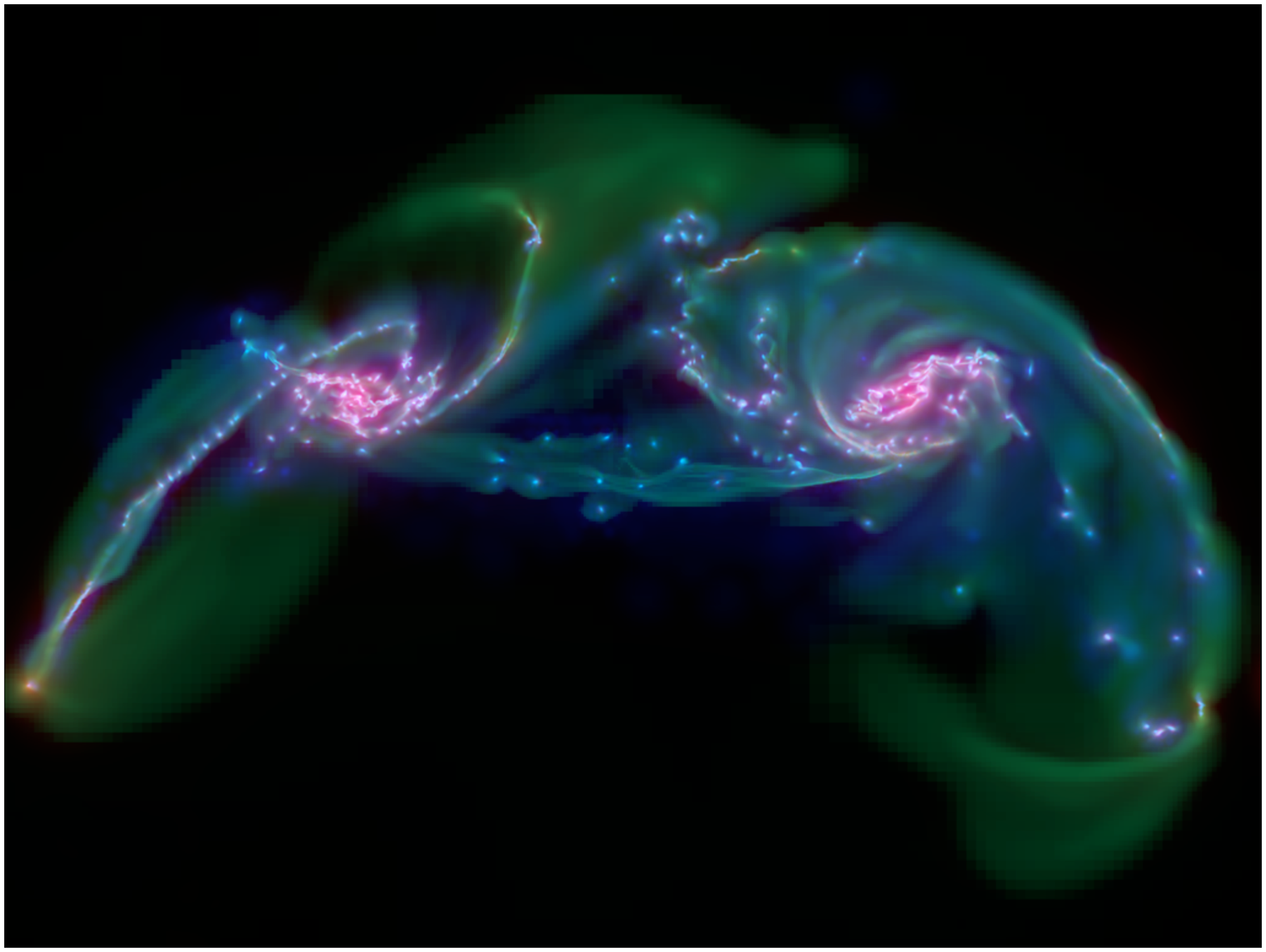}\\
\vspace{-0.34in}
\hspace{0.41in}
\includegraphics[angle=90, width=5.7in]{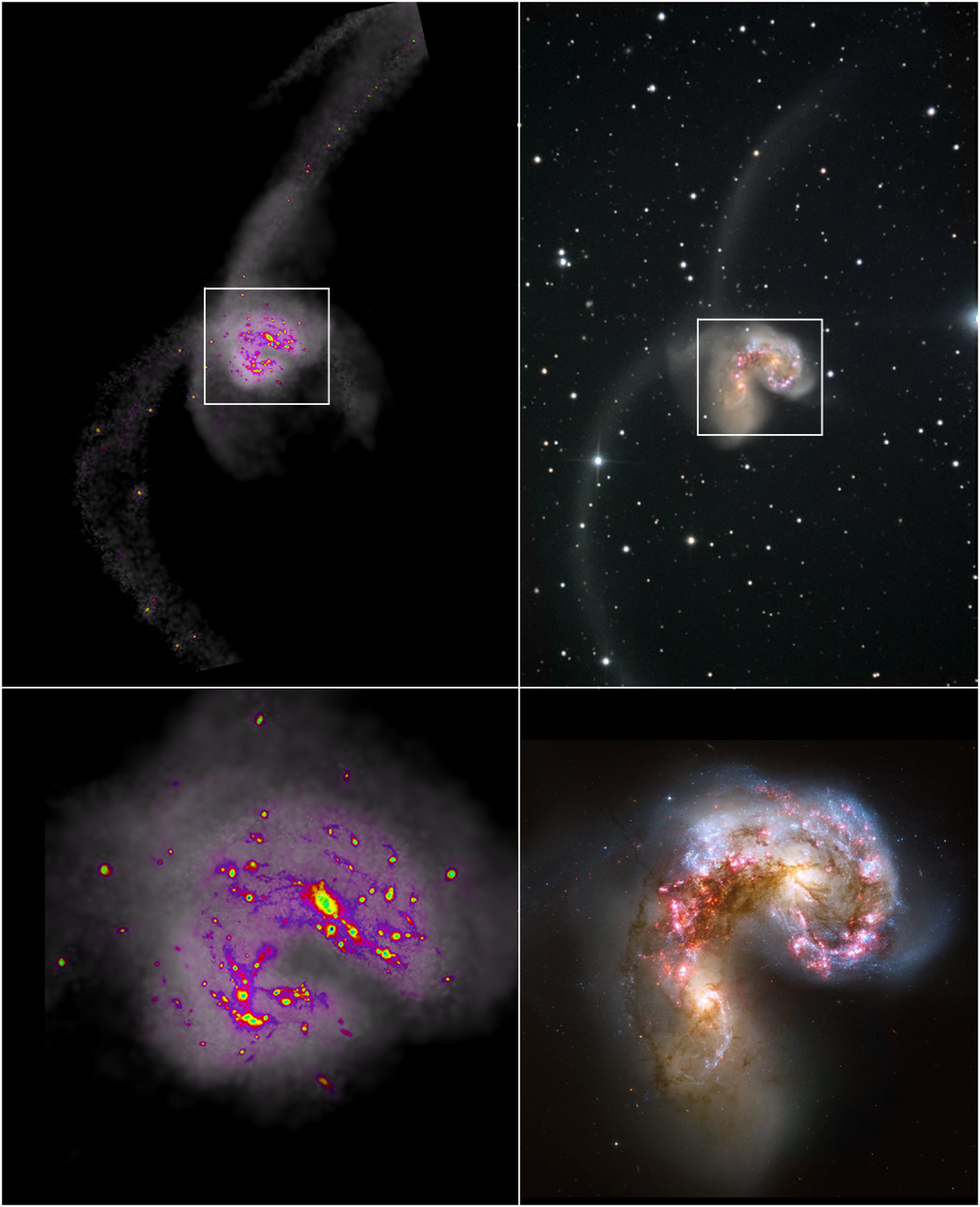}
\caption{Top: RGB rendering of the system shortly after the first pericenter passage (blue: young stars, red: old stars, green: gas, $t\simeq 350$~Myr). Bottom: Stellar column density map shown in log-scale at $t \simeq 475$~Myr and compared to the Antennae (HST images).}
\label{fig:antennae}
\end{figure*}

The global SFR in our model (like most others) can be computed directly by integrating the Schmidt law over the gas density PDF above the threshold $n_{*}$. The problem therefore boils down to predict the PDF evolution during the merger process. As shown by several authors \citep{Elmegreen:2002p2535, Elmegreen:2004p2540, Wada:2007p2426}, the multiphase structure of the ISM is built up from complicated processes, involving radiative losses (ultraviolet and infrared line cooling, molecular and dust cooling) as well as various heating mechanisms (cosmic rays and UV heating, supernovae and stellar feedback) and of course self-gravity. Surprisingly, numerical experiments have shown universal 
properties for the gas density PDF in isolated galaxies, with log-normal or power law distribution shapes. This was explained as a fundamental property of isothermal (or dissipative) self-gravitating turbulence. In this paper, our goal is to resolve this supersonic turbulence to compute self-consistently the density PDF and the resulting SFR in the course of a galaxy collision.

We use for that purpose a simple thermodynamical model mimicking gas cooling due to a detailed balance between atomic and fine structure cooling and UV radiation heating from a standard cosmic radiation background \citep{Haardt:1996p1167}. Assuming solar metallicity, we have computed the equilibrium temperature as a function of gas density to define our polytropic equation of State (EoS) $T=T_{\rm eq}(n_{\rm H})$. For densities between $n_{\rm H}=10^{-3}$~H/cc and $n_{\rm H}=0.3$~H/cc, our model has $T_{\rm eq} \simeq 10^4$~K, while above $n_{\rm H}=0.3$~H/cc, we have $T_{\rm eq} \simeq 10^4 \times (n_{\rm H}/0.3)^{-1/2}$~K. In this derivation, we neglected self-shielding of the radiation by the gas: gas cooling at very high density may have been underestimated, but we have also neglected the effect of local radiation sources such as OB stars as additional heating sources. Although our thermal model appears rather uncertain, it provides a reasonable route of gas dissipation, maintaining the gas temperature to a realistic average value at a given density. This EoS-based model produced a log-normal gas density PDF in isolated galaxies similar to the results of complete cooling/heating calculations (see section~\ref{sec:results}) and a realistic density power spectrum of ISM substructures \citep{bournaud2010}.

Another important ingredient is the thermal support added at small scales to avoid artificial fragmentation \citep{Truelove:1997p2564}. To ensure that the Jeans length is always sampled by at least 4 cells, we add an artificial pressure defined as $P_{\rm Jeans}=16\Delta x^2G \rho_{\rm gas}^2/\gamma \pi$. This technique, introduced in a different context by \cite{Machacek:2001p2609}, efficiently prevents the formation of spuriously fragmenting clumps in galaxy formation simulations \citep{Robertson:2008p2656, Agertz:2009p356}. 
From our EoS, we can compute the typical density at which this pressure floor dominates, namely $n_{\rm Jeans} \simeq 6 \times (\Delta x/100~{\rm pc})^{-4/3}$~H/cc and the equivalent gas temperature as $T_{\rm Jeans} \simeq 2500 \times (\Delta x/100~{\rm pc})^{2/3}$~K. This density corresponds to our minimum thermal Jeans mass, which defines 
our mass resolution $m_{\rm res}=M_{\rm Jeans}$.  

We use a quasi-Lagrangian refinement strategy: each cell for which the mass exceeds $m_{\rm res}$ is subdivided into 8 
children cell, down to the maximum level of refinement. In order to study the convergence properties of our system and identify qualitative changes, we 
performed a {\em low resolution} model with $\Delta x = 96$~pc, $m_{\rm res}=10^6$~M$_\odot$ and a {\em high resolution} one with $\Delta x = 12$~pc, $m_{\rm res}=4 \times 10^4$~M$_\odot$. Note that in the low resolution run, the gas cannot cool significantly below $10^4$~K, while in the high resolution case, it can reach a minimum temperature of $T_{\rm Jeans} \simeq 500$~K at the Jeans density.

We have fixed the star formation efficiency parameter to $\epsilon_*=0.01$, adjusting the star formation density threshold to $n_*=0.1$~H/cc (resp. $n_*=8$~H/cc) for the low (resp. high) resolution run. This ensures that both simulations initially have the same initial SFR of $\sim 1$~M$_\odot$/yr per galaxy in isolated disks, in agreement with the KS law of local spirals. Since in both cases, $n_*$ is significantly below $n_{\rm Jeans}$, the star forming part of the PDF is well sampled. 

\begin{figure}
\includegraphics[width=6in]{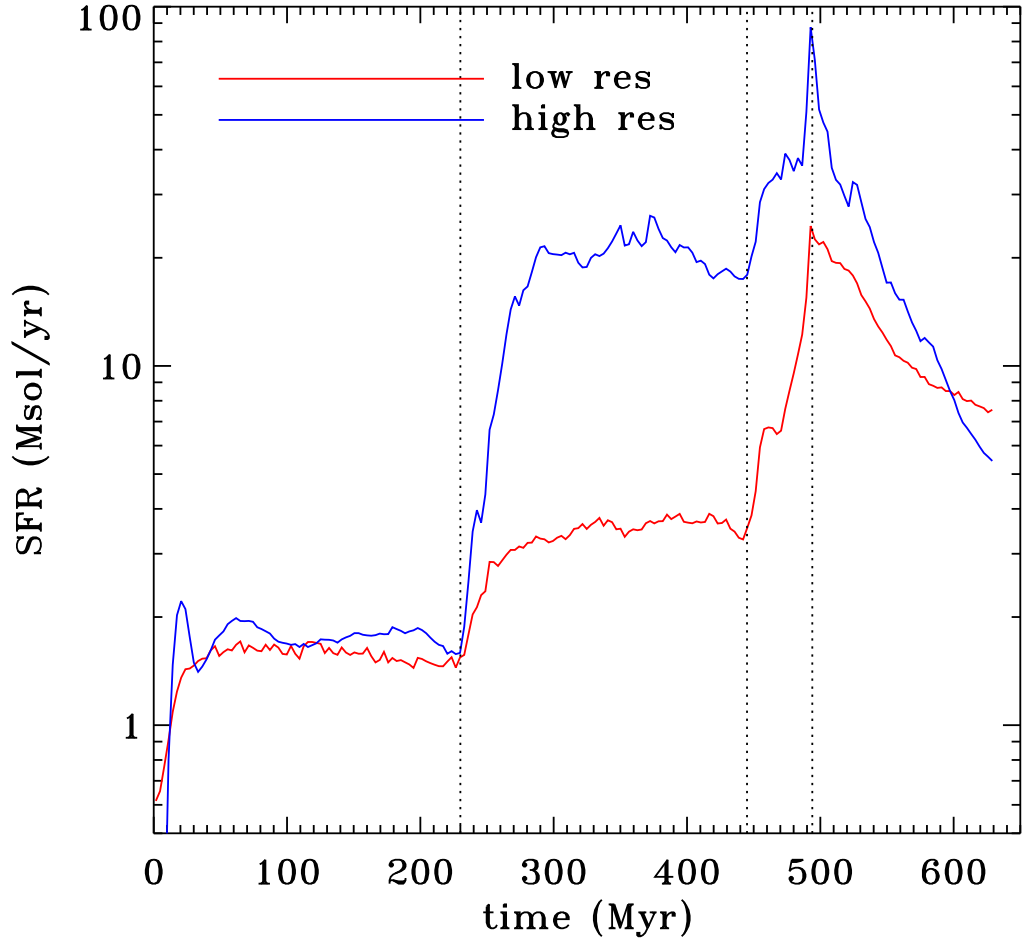}
\caption{Star formation history in our simulations for the low (red) and high (blue) resolution runs. Times for first pericenter, second pericenter and final merger are indicated by vertical dotted lines.}
\label{fig:sfr}
\end{figure}

\subsection{Merger parameters and initial conditions}

We model a pair of galaxies in a box of size 200~kpc, with isolated and outflow boundary conditions. Each galaxy is embedded in a live halo with a \cite{Hernquist:1993p3304} density profile with masses $M_{\rm h,1}= M_{\rm h,2}= 1.8 \times 10^{11}$ M$_\odot$ and scale lengths equal to the truncation radii $r_{\rm h,1}=r_{\rm h,2}=30.8$~kpc. The pre-existing stars are described as two exponential disks of masses  $M_{\rm *,1}=M_{\rm *,2}=3.6 \times 10^{10}$ M$_\odot$, scale length $r_{\rm d,1}=r_{\rm d,2}=4.4$~kpc and truncation radius $r_{\rm max,1}=22$~kpc, $r_{\rm max,2}=13.2$~kpc respectively. The disks scale height are $h_{\rm d}=0.2 r_{\rm d}$. A central bulge with B/D=1 is added, with a \cite{Hernquist:1990p3318} profile with $a=4.4$~kpc. The gas distribution follows the stellar disk profile with a total gas fraction of 10\% in each galaxy. The total number of dark matter particles was set to $N_{\rm dm}=8 \times 10^5$ and the initial number of stars to $N_{\rm old,*}=6 \times 10^5$ for both the low and high resolution simulations. 
We use the hyperbolic orbit proposed by \cite{Renaud:2008p2701} to reproduce the Antennae system. Our AMR grid has a coarse level $\ell_{\rm min}=7$ (i.e. $128^3$ cells) and a maximum level of refinement of $\ell_{\rm max}=11$ (resp. $\ell_{\rm max}=14$) for the low (resp. high) resolution simulation. 
 
\begin{figure}
\includegraphics[width=7in]{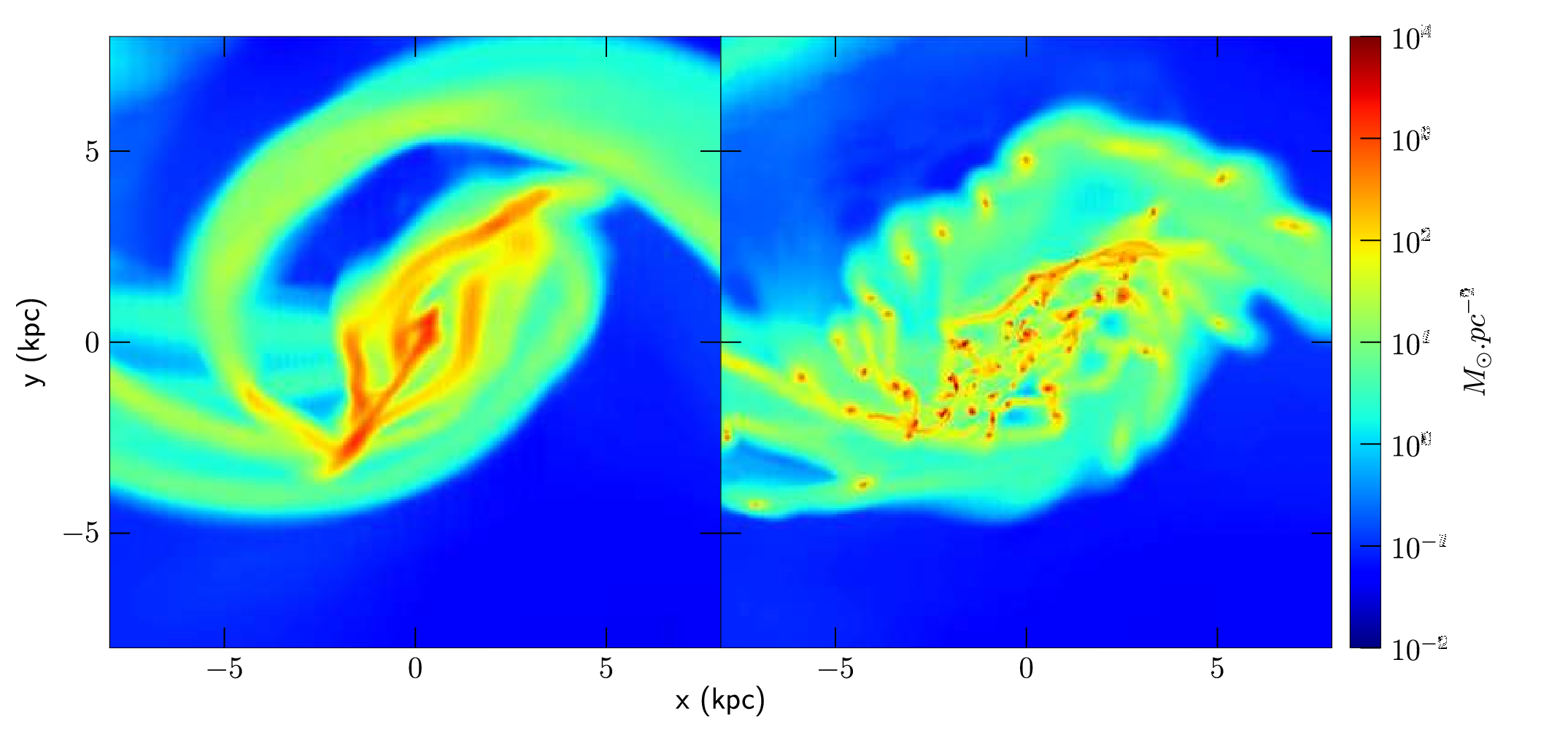}\\
\includegraphics[width=6in]{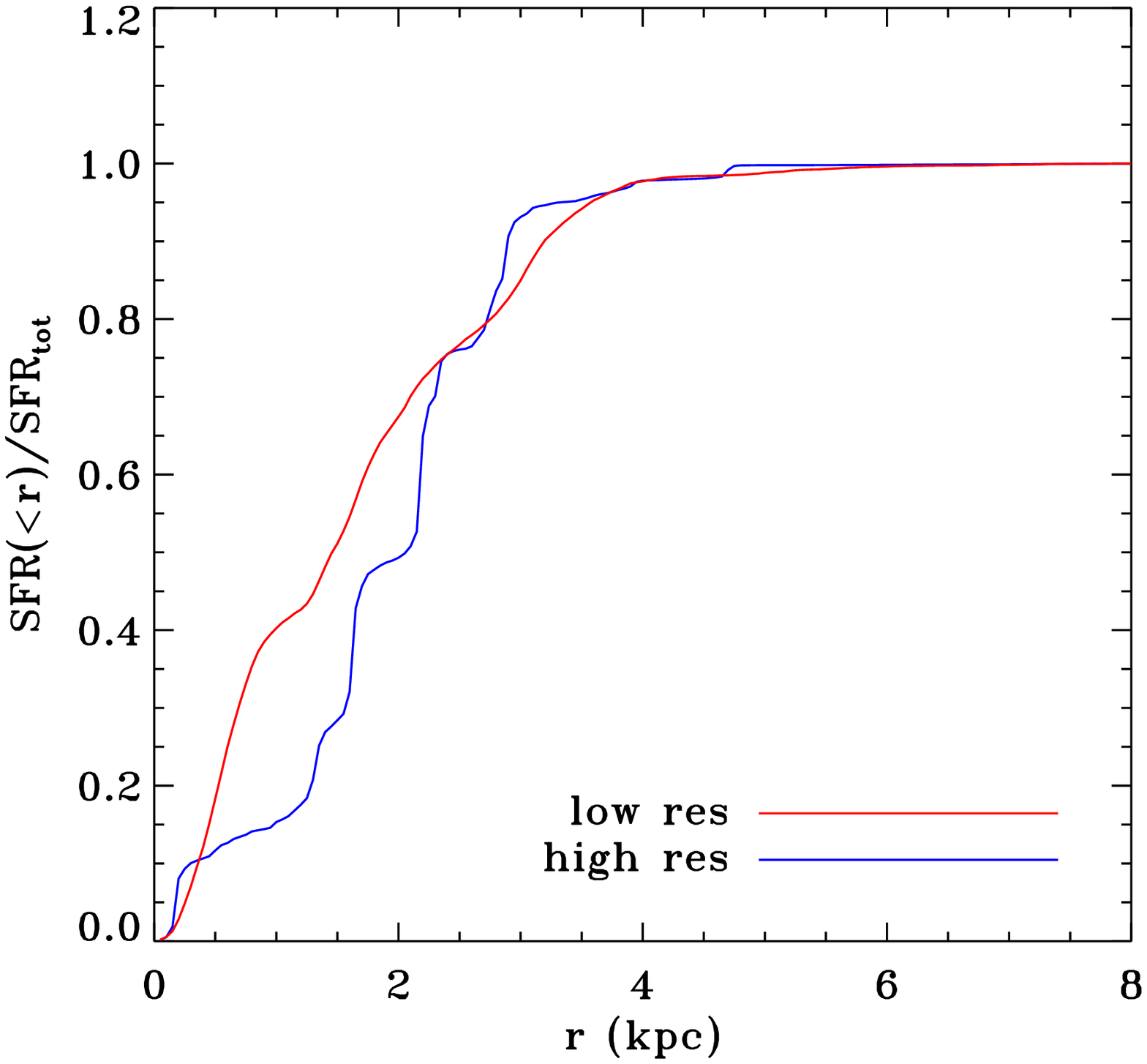}
\caption{Top panel: face-on view of the gas column density at time $t=350$~Myr for disk 1 for the low (left) and high (right) resolution runs. 
Bottom panel: corresponding fractional star formation rate as a function of radius. }
\label{fig:maps}
\label{fig:sfrprof}
\end{figure}

\section{ISM dynamics and star formation in a galaxy merger}
\label{sec:results}

\begin{figure}
\vspace{-1.1in}
\includegraphics[width=6in]{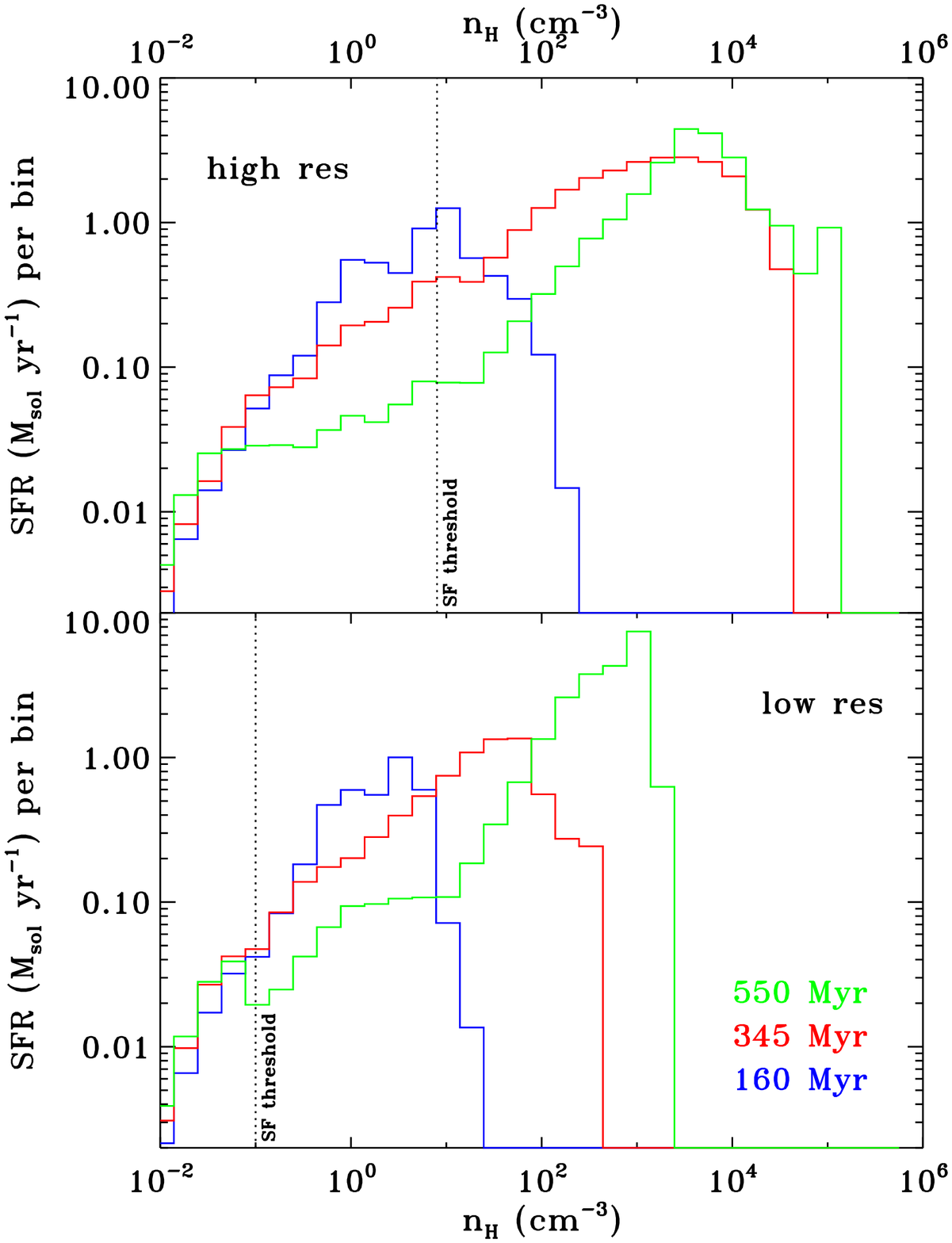}
\vspace{-1.5in}
\caption{Star-formation-weighted gas density PDF at three different epochs, before ($t$=160~Myr) and during the merger ($t$=345, 550~Myr), for the high (top panel) and low (bottom panel) resolution simulations. In each case, the star formation density threshold is indicated by the dotted vertical line. }
\label{fig:pdf}
\end{figure}

\begin{figure}
\includegraphics[width=6in]{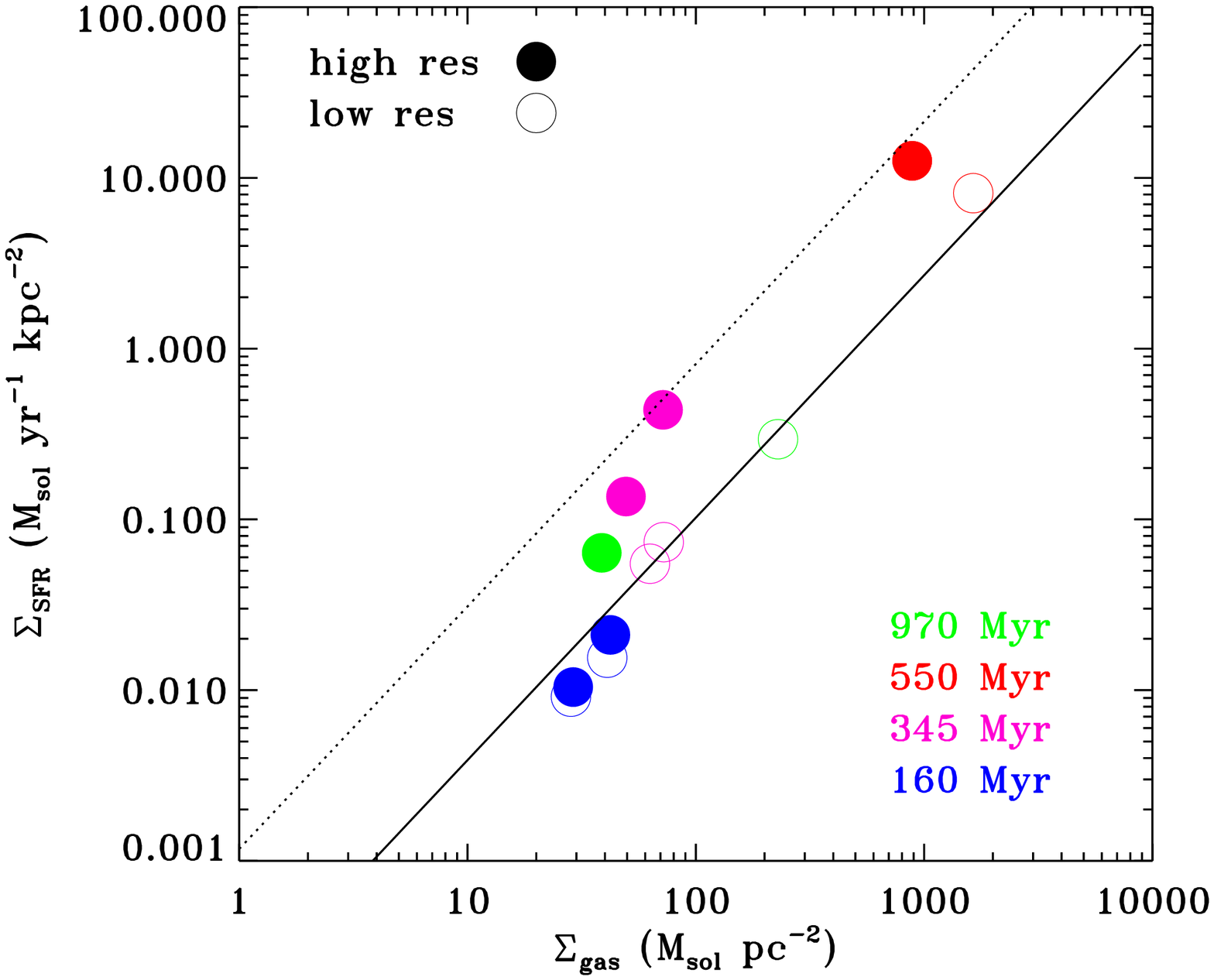}
\caption{Time evolution of the two merging galaxies in the KS diagram, comparing the low and high resolution runs. The solid line is a fit to the average KS law for quiescent galaxies, while the dotted line is the KS law for starburst galaxies \citep[from][]{Daddi:2010p4354}.}
\label{fig:ks}
\end{figure}

Figure~\ref{fig:sfr} shows the star formation history of our 
merger at low and high resolution. At first pericenter passage ($t\simeq 250$~Myr), the SFR rapidly rises in both runs but with a dramatic difference in amplitude. Around $t=450$~Myr (second pericenter passage) and $t=500$~Myr (final coalescence), the SFR increases again, now more significantly for the low resolution run, and it steadily declines after the merger. 
 
Although the pre-merger isolated disks have similar SFRs, the global SFE during the mergers is totally different depending on the resolution, i.e. depending on whether we model a smooth and warm ISM or a cloudy multiphase ISM at high resolution. SFRs are discrepant by a factor of $\sim$10 after the first pericenter and $\sim$5 in the final merger stages. This is due to a completely different ISM dynamics building up two different density PDFs between the low and the high resolution runs (Fig.~\ref{fig:pdf}) and different spatial distribution of the gas (Fig.~\ref{fig:maps}).

The low-resolution model, with a relatively smooth and warm ISM, follows the traditional mechanism of merger-induced starbursts, where tidal torques excite a strong $m=2$ mode and drive gas inflows below the corotation radius. The typical $m=2$ gas response and growing central concentration are seen on Figure~\ref{fig:maps}: the enhanced star formation takes place mostly in the central kpc (Fig.~\ref{fig:sfrprof}).

The gas evolution largely differs in the high resolution run with a cloudy multiphase ISM. An $m=2$ mode excited by tidal torquing is still visible, but the clumpy gas builds a much more modest central density peak than the low-resolution model: star formation is now spread over 2~kpc (Fig.~\ref{fig:sfrprof}).

The gas response to the interaction is actually dominated by fragmentation in many dense clouds scattered along the spiral arms, with a classical ``beads on a string'' morphology (see Fig.~\ref{fig:maps}), with typical clump masses of $10^{6-8}$~M$_\odot$. These high clump masses result from the gas turbulent motions, which are increased by the interaction, as already studied in \cite{bournaud08}. The gas velocity dispersion in our isolated disk at high resolution is typically 10-15~km~s$^{-1}$ but it increases by a factor of 2--3 in most regions, and locally by a factor of 5 or more, during the interaction. The Jeans length increases in similar proportions, so the typical mass of gas clouds formed by gravitational instabilities at fixed average density is increased by factors of typically 10 to 100. Such increased gas turbulence is observed in interacting galaxies \citep[e.g.,][]{dme95} and the associated formation of supermassive gas clouds studied by \citet{elm93}.

Using the theory of gravitational instability in a cylinder \citep{Chandrasekhar:1953p3341, Ostriker:1964p3346, Elmegreen:1979p3362}, we can compute the spiral arm stability criterion $q=\sigma^2/2G\mu$ where $\mu$ is the linear mass density along the arm. We found $\mu \simeq 4000$~M$_\odot$/pc on average in both simulations. At low resolution, the minimum sound speed always lies above 5~km/s, so that $q>1$ everywhere and the arm remains stable. At high resolution, the minimum sound speed can be lower (around 2.5~km/s) so that $q$ can be as low as 0.25 in quiescent regions, and the spiral arm can fragment into clumps in a free-fall time. As a consequence, the gas density PDF strongly evolves towards very high densities in the high-resolution model (see Fig.~\ref{fig:pdf}), and this evolution if achieved rapidly after first pericenter passage, and with little changes in the later stages: the star formation rate follows this evolution driven by ISM fragmentation in massive and dense clouds.

The low resolution model produces an artifically stable ISM and does not resolve this process. The gas response is dominated by the gradual inflow of gas towards the nuclear region with a timescale of a few $10^8$~yr (while dense gas clumps formed in a few $10^7$~yr). The associated density PDF increases slowly and continuously towards higher and higher densities as a result of the increasing gas concentration. This is why this model produces slower and more centrally-concentrated star formation than the high-resolution model. Note that our high resolution model do not resolve the clumpy ISM in the pre-merger isolated discs. This could affect our global SFR calibration in the initial conditions, but the initial gas density PDF would look very similar, except for the highest density tail. 

\section{Implications}

\subsection{The Antennae and other local mergers}

Although merger-induced star formation in our model is not primarily driven by an inflow of gas, and less concentrated than in earlier models, it remains relatively concentrated near the center of the merging systems (Fig.~3): there is still a tidally-induced inflow, and the central regions are denser and more prone to star-forming instabilities. This is consistent with star formation in ULIRGS being in general centrally concentrated (assuming ULIRGs are mergers). Nevertheless, our model also explains that the interaction-induced star formation can also be, for a part, radially extended (see Fig.~\ref{fig:sfrprof}). This can explain why a number of interacting galaxies actually show extended star formation with SSCs forming at several kpc from their center, such as Arp~140 \citep{Cullen:2006p4437} and the Antennae \citep{Wang:2004p4541}. 

As for the Antennae, the orbit of which is matched by our simulation, there is a general consensus that we are witnessing the merger close to the second pericenter passage, when the two disks are still well separated \citep{Renaud:2008p2701}. This corresponds to an epoch close to 450~Myr in our simulation (Fig.~\ref{fig:antennae}). Our low resolution models reaches SFRs around 10~M$_{\sun}$~yr$^{-1}$ just after the second pericenter passage and during only 50~Myr: these properties are in broad agreement with recent SPH simulations by Karl et al. (2010) and with the observed SFR (Zhang et al. 2001). Karl et al. (2010) then proposed that this high SFR and the extended SF in the Antennae results from the system being observed just at the particular instant of overlap between the two disks. Our high resolution model, however, shows that high SFR around 20~M$_{\sun}$~yr$^{-1}$ and relatively extended SF can be produced during a longer period (300~Myr) and does not require the system to be observed at a particular and brief instant.

\subsection{General star-formation laws and the starburst regime}

Observations suggest a dual law for star formation, where the integrated gas consumption timescale ($\Sigma_{SFR}/\Sigma_{gas}$) is relatively low for quiescent star-forming disks, and higher for starbursting ULIRGs and SMGs (likely major mergers), as pointed out independently by \cite{Daddi:2010p4354} and \cite{Genzel:2010p4738}. To compare our models with these observations, we retrieved integrated properties such as half-light radii, total gas mass and total SFR at several instants. The low-resolution model, where the starburst is driven only by the central gas inflow, does not show the observed change in $\Sigma_{SFR}/\Sigma_{gas}$: the SFR increases during the merger, but only in proportion corresponding to the increase in the global gas density $\Sigma_{gas}$, and this model remains close to the standard relation for isolated disks (Fig.~\ref{fig:ks}). The high-resolution model has its starburst driven mostly by increased gas turbulence and fragmentation. The gas density increases mostly on small scales in dense clumps throughout the system: this process does not affect the total effective size of the gas component, so the observable global density $\Sigma_{gas}$ has only a modest increase. At the same time, the starburst is even stronger than in the low-resolution model. This processes brings our model in agreement with the ``starburst vs. quiescent KS law'' pointed out by observations, throughout the duration of the merging process. Clustered star formation in high-resolution merger models can also affect the final structure of the resulting early-type galaxies (Bois et al. 2010).

We thus propose that these recent observations unveiled a ``starburst regime'' where the efficiency of star formation on small scales and at high densities is unchanged, but exacerbated gas turbulence and fragmentation into massive clouds result in faster gas consumption and higher integrated SF efficiency. The adopted SF law inside the clouds is not the key ingredient in the interpretation, the main effect being the rapid evolution of the density PDF. Different star formation models can indeed lead to different quantitative predictions, but the main qualitative change comes from resolving the high density tail of the PDF \citep{gov2010}. Previous models of galaxy mergers did not resolve ISM turbulence and clouds, and could not unveil the physical processes driving this starburst mode. The actual process of star formation in starbursting mergers cannot be captured with sub-grid models on scales larger than 100~pc, and requires clustered star formation in a multiphase ISM to be directly resolved.

\section*{Acknowledgments}
This work was granted access to the HPC resources of CINES and CCRT under the allocations 2009-SAP2191 and 2010-GEN2192 made by GENCI.




\label{lastpage}
\end{document}